# The Role of Data Analysis in the Development of Intelligent Energy Networks

Zhanyu Ma[1], Jiyang Xie[1], Hailong Li[2,3,*], Qie Sun[4,*], Zhongwei Si[5], Jianhua Zhang[6], Jun Guo[1]

[1.] Pattern Recognition and Intelligent Systems Lab., Beijing University of Posts and Telecommunications, Beijing, China.

[2.] School of Business, Society and Engineering, Mälardalen University, Västerås, Sweden

[3.] Tianjin Key Laboratory of Refrigeration Technology, School of Mechanical Engineering, Tianjin University of Commerce, Tianjin, China

[4.] Institute of Thermal Science and Technology, Shandong University, Ji'nan, China

[5.] Key Laboratory of Universal Wireless Communications, MOE, Beijing University of Posts and Telecommunications, Beijing, China

[6.] State Key Lab of Networking and Switching Technology, Beijing University of Posts and Telecommunications, Beijing, China

**Abstract**

Data analysis plays an important role in the development of intelligent energy networks (IENs). This article reviews and discusses the application of data analysis methods for energy big data. The installation of smart energy meters has provided a huge volume of data at different time resolutions, suggesting data analysis is required for clustering, demand forecasting, energy generation optimization, energy pricing, monitoring and diagnostics. The currently adopted data analysis technologies for IENs include pattern recognition, machine learning, data mining, statistics methods, etc. However, existing methods for data analysis cannot fully meet the requirements for processing the big data produced by the IENs and, therefore, more comprehensive data analysis methods are needed to handle the increasing amount of data and to mine more valuable information.

**Index Terms**

Intelligent energy networks (IENs), data analysis, smart grids, smart district heating networks, smart natural gas networks

## I. INTRODUCTION

The intelligent energy networks (IENs), which cover smart grids, smart district heating (DH) networks, and smart natural gas (NG) networks, can be defined as networks that intelligently optimize energy exchange based on bilaterally sharing information from both producers and consumers, and also refer to the integration of advanced information and communication technologies (ICT) with conventional energy networks [1]. In IENs, networking, ICT and data analysis methods are integrated into every aspect of the energy system including energy generation, energy transmission, energy distribution and consumer appliances [2]. The IENs have developed rapidly in recent years in order to meet the increasing demand for energy delivered in a robust, flexible, environmentally friendly and cost effective way [1]. The developmental stages of IENs are illustrated in Figure 1, which illustrates that the trend follows that of the development of smart meters. Smart meters are electronic devices that measure energy consumption and operate two-way communication regarding billing information and the status of energy systems. The capability of operating two-way communication is the most important feature that distinguishes smart meters from conventional meters. The evolution of smart meters has resulted in a rapid increase in data about energy systems. For example, the increase in meter reading frequency from once a month to every 15 minutes yields about 3,000 times more data. The large volume of data opens up new opportunities for a better understanding of consumer behaviour, clustering energy consumption patterns, and further optimizing energy production and distribution.

---

* Corresponding author.

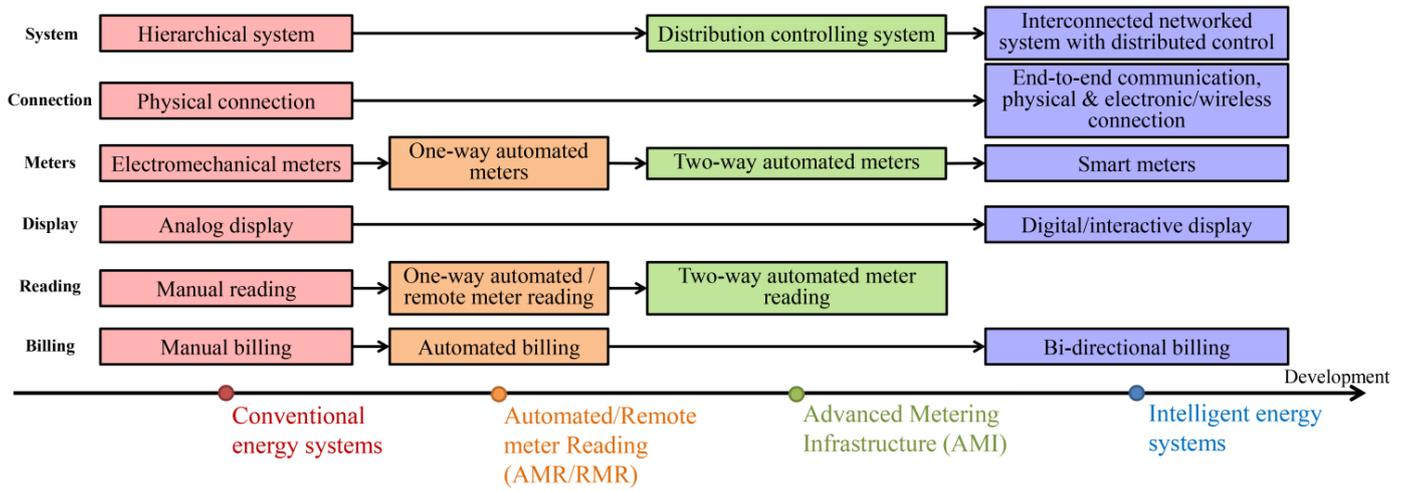

Fig. 1: Stages of development of an energy network.

With the rapid development of energy networks, data analysis becomes particularly important and essential to provide the foundation for IENs. Traditional energy networks can only collect local statistics from a small amount of data. However, they cannot efficiently handle the huge volume of data generated in IENs. Although manually collecting and analyzing data are acceptable in traditional energy networks, an advanced system dependent on modern computer technology can *not only* automatically collect data, process data, and deliver results, *but also* analyze the results against expected demand, standards, and concurrent data in IENs [3]. With the development of IENs, more and more data can be collected from energy networks, thus requiring the development of robust and efficient data analysis methods. The role of data analysis in IENs is illustrated in Figure 2. Data analysis can extract valuable information from a large amount of data, which can then be used for model design and algorithm implementation [2]. Many data analysis methods, which include pattern recognition and machine learning, have been developed and adopted mainly in three areas in IENs: i) malfunction diagnosis of energy networks to identify locations of current faults and to forecast locations of future faults, ii) data mining of consumer behaviours to cluster consumers, to provide suggestions in adjustment of users' individual consumption patterns, and to design policy for specific consumers, and iii) prediction of energy demand to provide information to support energy producers in the accurate planning for energy production, and to manage entire energy networks in a more efficient, precise and reliable way [4]. With the explosion in the growth of data volume, IENs have entered the era of energy big data. Energy big data of IENs exhibit not only volume, but also velocity and variety as the features which are "3Vs" model describing big data [2]. Smart energy meters are usually deployed at the scale of multimillion or multibillion units to collect real-time energy generation and consumption data in nodes at all levels of IENs. Many structured and unstructured data are collected from multiple sources and categories that are relevant to the energy generation process, energy planning process, energy distributed transmission and storage process and energy consumption process etc. In addition to the data generated by smart energy meters, other data such as weather and market information are collected to optimize the energy system. When more data is available, more valuable results can be obtained from data analysis. Accordingly, the requirements of data analysis keep increasing. Therefore, reliable and efficient data analysis methods, which are suitable for IENs, are necessary. Mining the big values from the energy big data in the field of energy can stimulate the development of IENs. However, existing methods of data analysis may not meet the requirements on energy big data and, therefore, more advanced data analysis methods are urgently needed to handle the increasing amount of data and to mine for more valuable information.

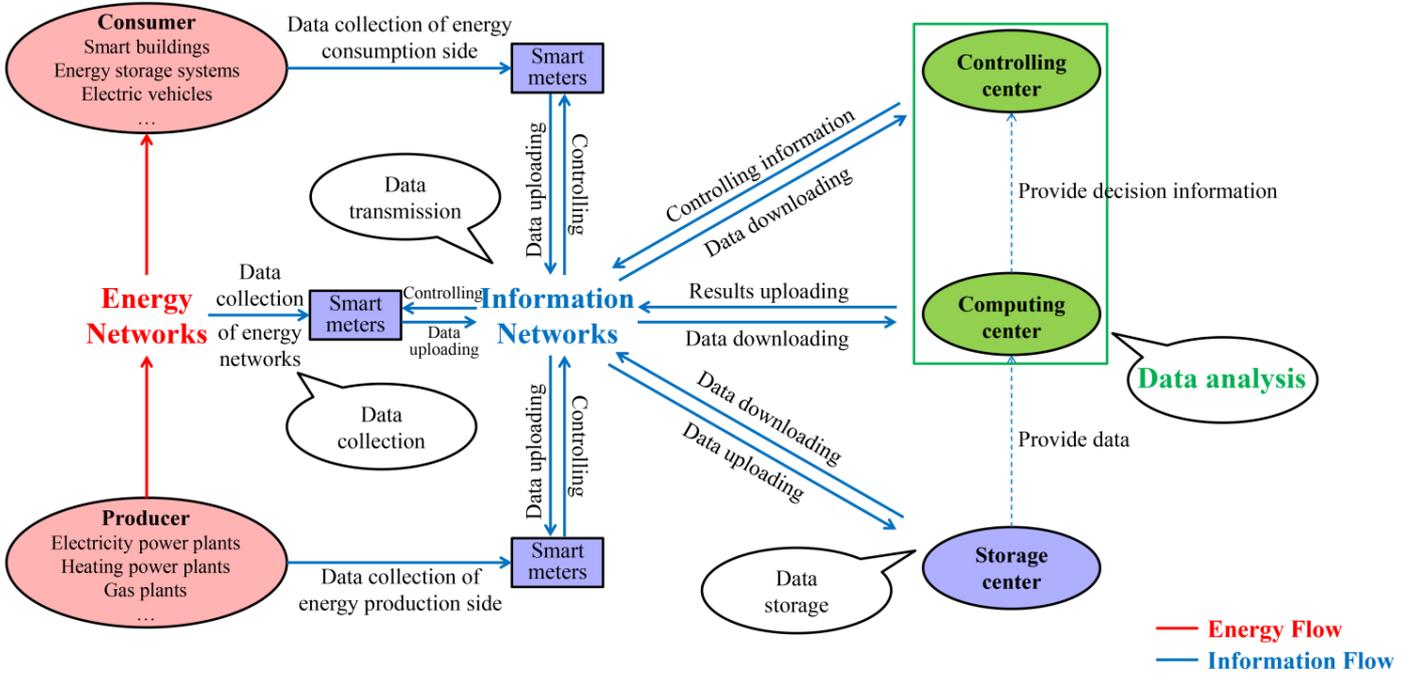

Fig. 2: Data collection, transmission, storage, and analysis in IENs.

The objective of this article is to identify the need for data analysis and to provide insights regarding the development of data analysis technologies through reviewing state-of-the-art applications of data analysis in the area of IENs. The remainder of the article is organized as follows: Section II introduces the development of energy meters and data collection. The typical applications and frequently adopted methods of data analysis are reviewed in Section III and Section IV, respectively. In section V, the article summarizes the main findings and provides suggestions on the development of future data analysis.

## II. ENERGY METERS AND DATA COLLECTION

*A. The Development of Smart Energy Meters and their Applications*

Smart energy meters, including smart electricity meters, smart heat meters, and smart gas meters, are the most fundamental components of the intelligent energy networks used to measure energy flows and exchange information on energy consumption and the status of energy networks between utility companies and consumers [1].

The most common type of electricity meters is the electro-mechanical meter. The advantage of electro-mechanical meters lies in the reliability of measurement by counting the revolutions of an electrically conductive metal disc. However, manually reading, electricity consumed, and creep phenomenon are difficult to avoid. Smart electricity meters, which are electronic and intelligent and have replaced traditional electro-mechanical meters in recent years, integrate two-way communication and other smart controlling functions.

Heat meters can be divided into dynamic meters (including impeller meters and turbine meters) and static meters (including magnetic induction meters and ultrasonic meters). Smart heat meters can measure heat flows, which are hot water or stream. Meanwhile, they can also communicate, save and process measured data.

Gas meters are usually used to measure the flow volume of gas and to adjust the measurement according to the temperature and pressure. The most common types of gas meters are diaphragm/bellows meters, which can be found in most residential and small commercial installations. The others types of gas meters, such as rotary meters, turbine gas meters, orifice gas meters, ultrasonic flow meters and coriolis meters, are also used in some special applications.

Before smart energy meters, conventional electro-mechanical meters were the dominant type of devices for electricity flows measurement. The measured data were usually displayed on a traditional counter, which had to be manually recorded in conventional energy systems. With the emergence of smart energy meters, more comprehensive functions have been developed and integrated such as regular and precise metering. This is the most fundamental function of an energy meter i.e. to gain information on both the energy supply and demand sides. Data recording is of great significance to enable utilities to monitor the status of energy networks. It is also useful for consumers to know details about their energy consumption. In addition to this, the alarm function can alert to an overload of energy consumption according to the data recording of meters. Smart energy meters should include a communication module to facilitate two-way communication between them and the service providers to transmit the measured data and the instructions for linked appliances. With the unprecedented rise in Machine-to-Machine (M2M) communications over wireless links, more wirelessly connected smart energy meters with robust and reliable technologies, such as ZigBee-based M2M communication, have been deployed and connected in a complex manner to promise higher efficiency and scalability in IENs [5]. Smart energy meters can influence consumer behaviour with regard to energy consumption through demand-side management. Energy systems also face problems of energy theft and system security. Identification of unauthorized consumption and encryption of information is key to solving these additional problems. The functions of smart energy meters are able to enhance the control over IENs and improve the stability of systems.

The applications of smart energy meters bring numerous benefits to IENs. These include better access for load management, more accurate billing, data quality and safety improvement, environmental benefits improvement, and detection of interruptions and energy theft. The evidence from smart grids would suggest that systems' stability and security is of essential importance to the development of IENs.

With the deployment of smart meters, more data about consumers' energy consumption are available. China is leading the way, with more than 435 million devices installed, followed by the United States. Member states of the EU are required to roll out smart metering systems with a target of reaching 80% of consumers by 2020. This would equate to the installation of over 270 million smart electricity meters. Global smart meter installations are predicted to hit about 800 million by 2020 [6]. The deployment of smart heat meters is far behind that of smart electricity meters. Europe has been the largest market for smart heat meters to date, with approximately 6,000 DH networks. China is another big market for smart heat meters. If every apartment or house was equipped with a heat meter, the total number would reach around 150 million. In relation to gas meters, at present there are a total of 400 million mechanical meters in operation worldwide, and smart meters are gradually replacing them. It is estimated that about 28.4 million smart gas meters have be deployed in the period 2010-2016.

Figure 3 compares different types of smart meters with the corresponding deployments in a range of countries. The bar charts represent the number of smart electricity, heat, and gas meters which have been deployed. It should be noted that the length of bar is proportional to the logarithm based 10 of the number of meters. As Figure 3 illustrates, the deployment of smart heat and gas meters is far behind that of smart electricity meters.

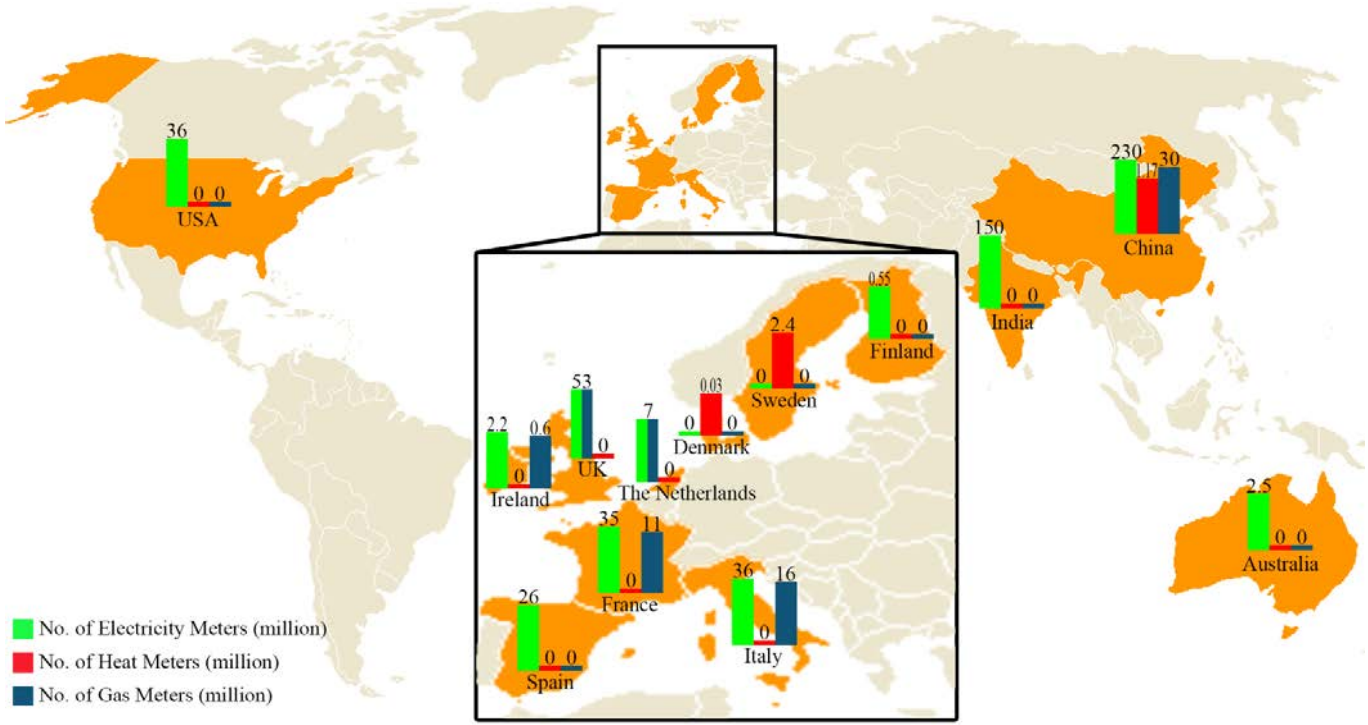

Fig. 3: Deployment of smart energy meters in a range of countries [1].

*B. The Available Data at Different Resolutions and Volumes*

The current generation of IENs usually requires smart meters to operate metering hourly or even more frequently. For example, the American utility PG&E offers metering systems that read consumption data every hour by default[*], and the electricity meters provided by British Gas for residential use operate on a half-hourly basis[†]. There are also examples that measure electricity usage in 15-minute intervals, e.g. the electricity meters provided by the CenterPoint Energy[‡]. The development of smart meters has enabled higher temporal resolution. For electricity meters, it is technically possible to measure at an interval of 1 minute, 1 second or even a millisecond. Assuming a frequency of 1 minute will give a volume of $4.2 \times 10^{14}$ data/year.

It can, however, be easier to get more information, such as load patterns and similarities, at a lower temporal resolution. One major issue with high temporal resolution is that the volume of data becomes large and these high volumes mean a high dimensionality of the clustering problem, which makes the algorithm computationally slower. The data can be aggregated, which although simple to implement, risks a loss of detail . Until now, little attention has been paid to the effects of data granularity on the performance and efficacy of statistical methods for analyzing energy demand profiles. Granell *et al.* [7] studied how temporal resolution of power demand profiles affects the quality and efficiency of the clustering process, and the consistency of cluster membership (profiles exhibiting similar behaviour). It was found that a band of temporal resolutions between 4-60 min showed better performance than either of the extreme resolutions. It was concluded that 8-min data would provide a useful basis; while data collected at frequencies slower than 30 min is not sufficiently reliable for most analytical purposes.

III. APPLICATIONS OF DATA ANALYSIS IN THE DEVELOPMENT OF INTELLIGENT ENERGY NETWORKS

Data analysis can extract useful information by scanning large amounts of data and automatically reporting interesting patterns [8].

*A. Clustering*

Removing energy peaks reduces the cost for both the grid owner and the consumer. In order to even out energy peaks, it is important to understand the behaviours of users and identify similarities between groups of customers. Additionally, understanding energy consumption patterns is fundamental to the optimization of resources. Data analysis can facilitate customer clustering based on consumption patterns which can help distinguish user types for demand-side management, tariff designing and switching, fault and fraud detection, and energy efficiency measurement [9]. Clustering can also be used to improve the predictive accuracy of energy models.

Chicco [10] suggested a four-step load pattern categorization procedure: i) data gathering and processing, ii) phase pre-clustering, iii) phase clustering, and iv) phase post-clustering. A number of clustering methods have emerged and they can be broadly classified into four categories: partitioning methods, hierarchical methods, density-based methods and grid based methods. Other clustering techniques that do not fit in these categories have also been developed, such as fuzzy clustering, artificial neural networks and generic algorithms.

*B. Demand Response*

Energy consumption is affected by the infrastructure, weather conditions and energy price but is also highly dependent on consumer behaviour, which is characterized by randomness, heterogeneous and interdependence. It is therefore very complicated to forecast the overall energy demand by simulating each consumer's behaviour using a conventional system based on a dynamic energy model. As an alternative, the application of data analysis technologies can provide a useful way to take consumer behaviour into consideration.

Furthermore, a better understanding of consumer behaviour and response to various changes in energy systems can be help in forecasting. Studies have shown there is a large potential in energy savings, which can be effected through influencing consumer behaviour [11]. With the development of IENs, an enormous amount of data about consumers' energy consumption can be recorded. This will enable a better understanding of energy systems and assist in the development of IENs. In future, energy systems could be designed in a more flexible way to incorporate fluctuations in demand and therefore allow for presumes and distributed renewable energy.

*C. Energy Production Optimization*

---

[*] http://www.pge.com/en/mybusiness/services/smartmeter/index.page
[†] http://www.britishgas.co.uk/smarter-living/control-energy/smart-meters/what-are-smart-meters.html
[‡] http://www.oncor.com/EN/Pages/Residential.aspx

An important way of saving energy is to optimize energy production according to the actual energy demand. The "Earth Hour" event sponsored by WWF offers a good example of how energy big data analysis can contribute to optimizing energy production at a specific point in time. "Earth Hour" suggests that consumers should turn off their unnecessary lights and electric devices at 20:30 on the last Saturday in March. The event has grown in popularity in many countries. However, the event itself will not contribute to saving energy if electricity producers continue to generate electricity to meet the usual demand for power at that time. To save energy, the electricity producers have to, in advance, know how many consumers will participate in the event and how much less electricity is likely to be used, and then adjust their production accordingly. In this situation, the technology of energy big data analysis could provide energy producers with information on how much energy is likely to be needed ahead of the event.

There is an even larger potential in energy saving by levelling out the energy load, i.e. peak clipping and valley filling. With smaller fluctuations in energy consumption, energy producers can employ a larger capacity to meet the base load while reducing the capacity for peak load, which is usually based on more expensive technologies. To this end, consumers must be involved in the reform of the entire energy system to reduce the peak-to-average ratio (PAR) by simultaneously optimizing users' energy schedules and lowering the overall energy consumption in the IENs, such as in an optimized game-theoretic demand side management scheme [12]. More importantly, this aim will not be possible without sufficient knowledge of demand responses (for more discussion on demand response, please refer to the last section).

*D. Energy Pricing*

Many different issues, especially energy price, can affect consumers' energy consumption patterns. Energy price is also one of the most important policy instruments that can be employed to reform energy systems. However, different consumers, typically characterized by age, income level, education and cultural background, may have different responses to the same changes in energy price. Energy big data analysis offers an efficient way to identify consumer's likely reactions to different energy prices.

Currently, there are many different types of pricing mechanisms in energy markets, e.g. constant pricing, spot pricing, tiered pricing, peak-valley pricing and time-of-use pricing. There is no "the best" mechanism suitable for all types of energy in different economic and social systems. It is often necessary to combine several types of pricing mechanisms to optimise efficiency in energy systems. Energy big data analysis technologies are highly effective in recognising consumers' attitudes towards different energy prices and subsequently estimating the impact of pricing mechanisms on energy production.

*E. Monitoring and Diagnostics*

A large volume of data can increase the reliability of power generation. For example, when a base load power plant goes down due to a maintenance issue, that generation capacity must be replaced with an alternative source of electricity, which can often be more expensive. Increasing the length of time between scheduled service events and spotting issues before they require a plant to shut down is beneficial to power plant operators and those relying on that electricity. To enhance the collaboration between engineers and customers, GE has developed a method based on energy big data, which are collected from its global fleet of gas-fired turbines on a daily basis[*]. Progress has been made in reducing the amount of time required to fix problems when they do arise. Another example is that with the increasing rate of installation of wind turbines, a huge amount of operating data becomes available. Data analysis technologies can be used to analyze the operation of wind turbines and develop target values, which enables the evaluation and detection of faults instantly and remotely[†].

## IV. DATA ANALYSIS METHODS IN THE DEVELOPMENT OF INTELLIGENT ENERGY NETWORKS

*A. Development of Data Analysis Methods*

Smart energy meters have been replacing conventional meters, providing energy consumption data to meter data management systems (MDMS) [4]. Integrating demand response requires the capability to coordinate the communication of demand-response events, and process data such as analysis of the predictive peak load's levels and the capacity available for each demand response [11]. Data analysis is needed to plan energy distribution, formulate strategies, and recommend energy policies. The need is significant and challenging *not only* in developing countries where required data, suitable methods and essential institutions are deficient, *but also* in industrialized countries although these limitations are less serious. A great deal of analytics algorithms and applications have been put forward focusing on data analysis to predict energy consumption and demand, to propose accurate energy consumption plans, to provide individual feedback to consumers on adjusting habits and reducing bills, and to design targeted projects for specific clusters of consumers [4].

---
[*] http://breakingenergy.com/2015/03/18/big-data-increasing-power-generation-reliability-and-saving-money-ge-says/
[†] http://www.siemens.com/innovation/en/home/pictures-of-the-future/digitalization-and-software/from-big-data-to-smart-data-remote-service-wind-turbines.html

Generally speaking, data analysis methods, including pattern recognition, machine learning, artificial intelligence, data mining and statistics methods, can be classified into two broad categories: supervised learning and unsupervised learning. These methods can be used in clustering, feature extraction/selection, classification and prediction. The most popular pattern recognition methods for data analysis in IENs are linear regression (LR), support vector machine (SVM) and neural networks.

Linear regression (LR) is a technique based on statistical learning and is used for analyzing the numerical data of a linear combination model with parameters including simple regression model (SRM), multiple linear regression (MLR), autoregressive integrated moving average (ARIMA), etc. LR has been used to predict energy demand and for benchmarking and energy mapping with linear regression models, especially in smart grids [13]. In practice, the performance of a regression model depends on the fitting approach adopted, while the best fit is typically evaluated by using the least squares method [14].

Support vector machine (SVM), which is a form of supervised learning machine algorithm, is a non-parametric statistical model based on the structural risk minimization (SRM) principle that offers great predictive capability for a limited sample size. SVM is a useful technique for data clustering since it is considered easier to use than neural networks while it is also used for load prediction and benchmarking in energy management systems [13].

Neural networks, which model arbitrary nonlinear relationships to any degree of accuracy by adjusting the network parameters consisting of artificial neural network (ANN), recurrent neural networks, feed-forward back propagation neural networks and redial basis function networks, and can be used as an analytical approach for energy demand prediction while offering certain advantages, such as no requirement of knowledge about internal system parameters and effective solution for multi-variable problems calculation [15]. Neural networks, with a back-propagation algorithm to increase precision, have been also used in benchmarking, user clustering, and fault diagnosis in electric or heat power systems.

Table I shows the application areas of the aforementioned data analysis methods. Depending on the application, the data analysis methods need to be adjusted in order to fit the features of data generated in IENs, for example, regulating parameters, integrating different methods and improving the methods. Adjusted data analysis methods are required to fill the lack of intuitive physical or other knowledge area interpretation of the IENs and to make a sound decision based on the methods [2].

TABLE I: Applications of data analysis methods for different application purposes.

| Methods | Applications | | | | |
| --- | --- | --- | --- | --- | --- |
| | Clustering | Demand Response | Energy Production Optimization | Energy Pricing | Monitoring and Diagnostics |
| Linear Regression | | √ | √ | √ | |
| Support Vector Machine (SVM) | | √ | √ | √ | √ |
| Neural Networks | √ | √ | √ | √ | √ |
| K-Means | √ | | | | |
| Kalman filter | | √ | √ | √ | |
| Gaussian Process | √ | √ | √ | √ | |
| Principal Component Analysis (PCA) / Independent Component Analysis (ICA) / Nonnegative Matrix Factorization (NMF) | √ | | | | |
| Learning Vector Quantization (LVQ) | √ | | | | |

*B. Two Data Analysis Method Application Examples*

With IENs becoming increasingly complex, the effectiveness of data analysis methods needs to be examined. In the following section, two examples of the application of data analysis in IENs are presented as a snapshot of this rapidly developing interdisciplinary research area.

*a. Prediction of heat demand in district heating systems*

In [16], a heat demand analysis model that adopts the Gaussian process (GP) technique has been proposed to predict the heat demand in district heating system (DHS) and to calculate the energy saving potential. The model utilized the heat demand data, the outdoor temperature data, and other factors in DHS to yield high prediction accuracy, where the mean absolute percentage error (MAPE) of heat demand prediction is between 5%-9%. Moreover, the proposed predictive strategy can be applied to energy consumption reduction in DHS. Stabilizing the return water temperature at a relatively low level means that pump usage can be reduced thus saving a large amount of electricity consumption. Figure 4 illustrates the comparisons of the measured and the predicted heat demand and the prediction error deviation distributions. The pump electricity saving (in $kWh/m^2/Hr$) in a commercial building (CB), an office building (OB), and an apartment building (AB) are $1.8\times10^{-6}$, $2.8\times10^{-6}$, $4.9\times10^{-7}$, respectively.

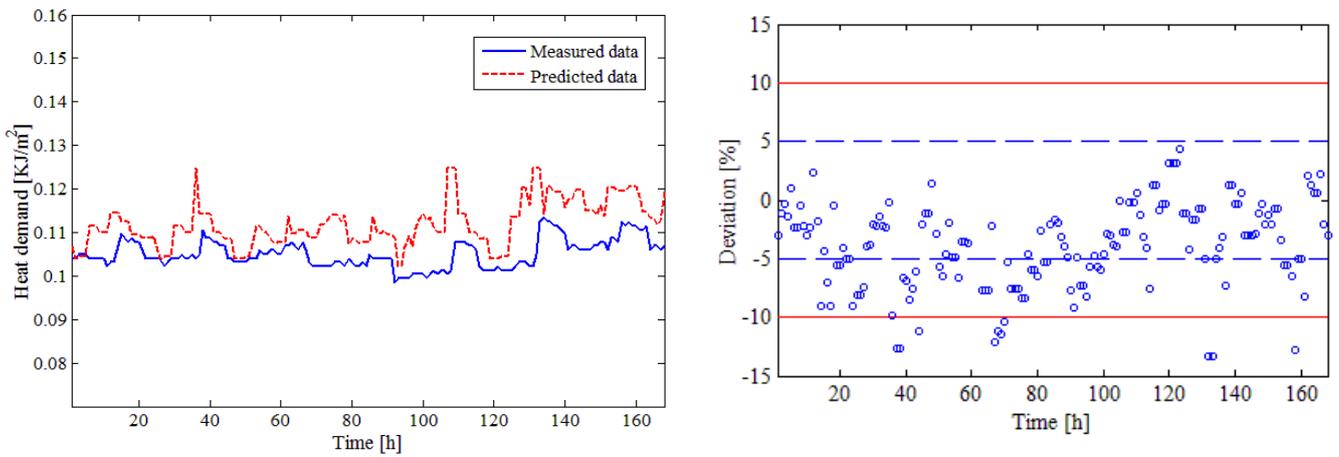

(a) Comparisons of predicted and measured heat demand in AB.    (b) Prediction error deviation distributions for OB.

Fig. 4: Heat demand prediction and energy saving with GP model [16].

*b. Analysis of key factors in heat demand prediction*

Deep neural network (DNN) is a popular data analysis method. It has various applications in IENs. In [17], a model based on Elman neural network (ENN) was employed to investigate and to analyze the impacts of the temperature, the direct solar radiance, and the wind speed on the heat demand prediction. To compare how different factors influence heat demand prediction, four models, namely ENN-A, ENN-B, ENN-C, and ENN-D, were constructed. ENN-A considered the temperature only. ENN-B considered both the temperature and the direct solar radiance. ENN-C considered the temperature and the wind speed. ENN-D considered the temperature, the direct solar radiance, and the wind speed, including all the three factors. Figure 5 shows the predicted and measured results (in MW) of dynamic heat demand and the distribution of percentage errors. The mean absolute percentage error (MAPE) for ENN-A, ENN-B, ENN-C, and ENN-D are 5.95%, 5.18%, 3.32%, and 4.37%, respectively. These experimental results show that the introduction of wind speed and direct solar radiation has opposite impacts on the performance of ENN and the inclusion of wind speed can improve the prediction accuracies. However, ENN cannot benefit from the introduction of both wind speed and direct solar radiation simultaneously [17]. Therefore, a more efficient network is required to handle this issue.

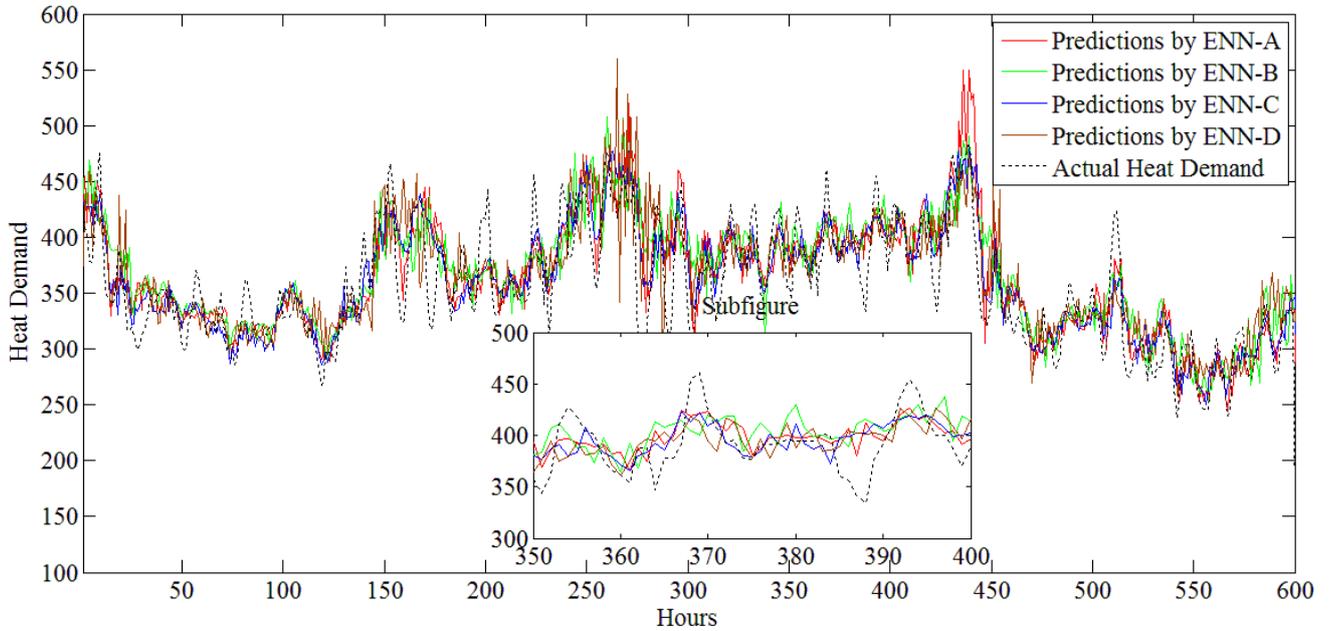

(a) Predicted and measured heat demands.

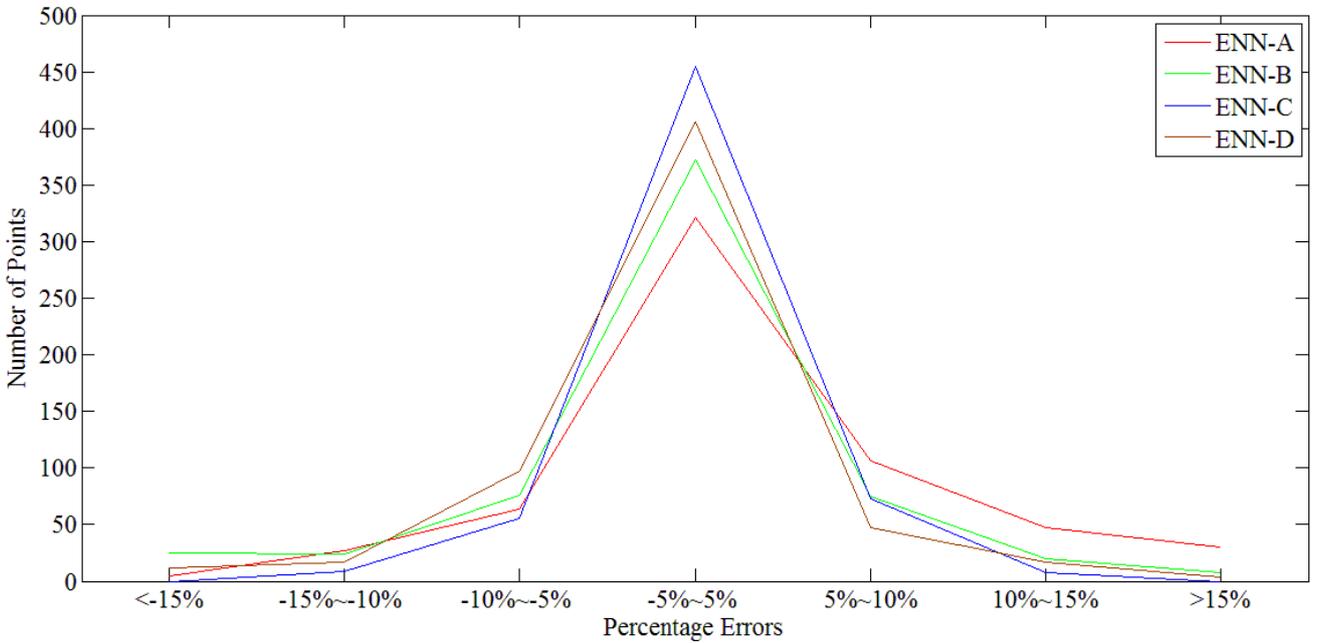

(b) Prediction error distribution.

Fig. 5: Key factor analysis in heat demand prediction with ENN [17].

*C. Challenges of Data Analysis Methods*

Although the methods of data analysis listed in Table I have been widely and successfully applied to, for example, clustering customers, predicting demands, optimizing production, pricing energy etc., some challenges and knowledge gaps still exist. For example, *over-fitting*, which yields poor predictive performance when the model overreacts to minor fluctuations in the training data, is a key and general problem in linear regression models. As the number of neural network layers increase, *gradient vanish*, where the obvious decreasing gradient of the hidden layer nodes makes the training accuracy rapidly decreased, is a common problem in deep neural networks. In the meantime, larger amounts of data from IENs is becoming available in the era of energy big data, which is characterized as huge data volume, numerous data variety, low value density, and high velocity [2]. The main challenges facing big data analysis are as follows: firstly, *not only* high-performance *but also* cost-effective storage and analysis methods for analyzing large-scale data are required; secondly, although more valuable information can be mined from a larger amount of data than before, the problem of low value density remains a the key hurdle; thirdly, new local data computing paradigms such as MapReduce and Spark bring great changes in data analysis methods and algorithm programming and implementing. Therefore, it is necessary to parallelize the existing traditional methods to adapt to the needs of energy big data analysis. In order to meet the aforementioned challenges, the following topics are important in future work: i) the development of scalable and interoperable computing infrastructures, ii) real-time and intelligent energy big data decision making based on historical data, iii) energy big data knowledge representation and processing by identifying the natures of the energy big data, and iv) new advanced methods suitable for parallel computation of large scale energy data [2].

## V. CONCLUSIONS

In this article, the role of data analysis in intelligent energy networks (IENs) has been reviewed and discussed. The development of IENs requires advanced data analysis technologies, which are significant in improving the performance of IENs. The applications of data analysis focuses on customer clustering, demand forecasting, energy generation optimization, energy pricing, monitoring and diagnostics. Correspondingly, typical methods, such as linear regression/prediction, support vector machine and neural networks, have been widely applied. In order to better meet the analysis requirements of energy big data, it is of great importance to develop comprehensive methods to effectively assist in the development of IENs.